# Ferroelectricity induced by interatomic magnetic exchange interaction

Xiangang Wan,[1,†] Hang-Chen Ding,[2] Sergey Y. Savrasov[3] and Chun-Gang Duan[2,4,‡]

[1]Department of Physics and National Laboratory of Solid State Microstructures, Nanjing University, Nanjing 210093, China
[2]Key Laboratory of Polar Materials and Devices, Ministry of Education, East China Normal University, Shanghai 200062, China
[3]Department of Physics, University of California, Davis, One Shields Avenue, Davis, CA 95616, USA
[4]National Laboratory for Infrared Physics, Chinese Academy of Sciences, Shanghai 200083, China

[†] Electronic address: xgwan@nju.edu.cn
[‡] Electronic address: wxbdcg@gmail.com or cgduan@clpm.ecnu.edu.cn

Multiferroics, where two or more ferroic order parameters coexist, is one of the hottest fields in condensed matter physics and materials science[1-9]. However, the coexistence of magnetism and conventional ferroelectricity is physically unfavoured[10]. Recently several remedies have been proposed, *e.g.*, improper ferroelectricity induced by specific magnetic[6] or charge orders[2]. Guiding by these theories, currently most research is focused on frustrated magnets, which usually have complicated magnetic structure and low magnetic ordering temperature, consequently far from the practical application. Simple collinear magnets, which can have high magnetic transition temperature, have never been considered seriously as the candidates for multiferroics. Here, we argue that actually simple interatomic magnetic exchange interaction already contains a driving force for ferroelectricity, thus providing a new microscopic mechanism for the coexistence and strong coupling between ferroelectricity and magnetism. We demonstrate this mechanism by showing that even the simplest antiferromagnetic (AFM) insulator MnO, can display a magnetically induced ferroelectricity under a biaxial strain.

The combination of different ferroic properties, especially ferroelectricity and (anti)ferromagnetism, provides additional degree of freedom to control magnetic and dielectric properties of the material. Such functionality is of great potential in applications to next-generation fast, portable and low-energy consumption data storage and processing devices. Unfortunately, magnetism and ferroelectricity tend to be mutually exclusive, as conventional ferroelectric perovskite oxides usually require transition metal (TM) ions with a formal configuration $d^0$, whereas magnetism, in contrast, needs TM ions with partially filled $d$ shells[10]. As a consequence, simultaneous occurrence of ferromagnetism and ferroelectricity is hard to be achieved, especially at room temperature. There are indeed some exceptions in Bismuth related magnetic oxides, *e.g.*, $BiFeO_3$ and $BiMnO_3$. In these compounds, however, the magnetism and ferroelectricity are widely believed to have different origins, rendering that the magnetoelectric coupling rather weak[9].

To gain a strong magnetoelectric coupling, vast efforts have been devoted to search the improper ferroelectricity where electric dipoles are induced by magnetism[6]. Phenomenological theory suggests that spatial variation of magnetization is essential for the magnetically induced electric polarization[11]. Several microscopic mechanisms[12-14], all emphasizing the importance of spin-orbital coupling, have also been proposed to explain the ferroelectricity in magnetic spiral structures. Electric polarization can also be induced by collinear spin order in the frustrated magnet with several species of magnetic ions[6,15-17]. There are also proposals to realize multiferroic state in composite systems or materials with nanoscale inhomogeneity[18,19]. Nevertheless, currently all of the known magnetically driven single-phase multiferroics require either Dzyaloshinskii-Moriya interaction (DMI)[20,21], which is small in strength, or competing exchange interactions in real space. Therefore, they generally have complex magnetic order, low transition temperature (below several ten K) and small electric polarization (generally two to three orders of magnitude smaller than typical ferroelectrics), making them still far away from practical applications. Therefore searching new mechanism for multiferroicity is of both fundamental and technological interest.

In this Letter, we demonstrate that regardless the spin-orbit coupling, simple interatomic magnetic exchange interaction already provides a strong driving force to break the inversion symmetry of the system, which is necessary for the occurrence of ferroelectricity. It is therefore a new microscopic mechanism for the coexistence and strong coupling between ferroelectricity and magnetism, and all of the magnetic insulators with simple magnetic structure need to be revisited. Using band structure calculations, we numerically confirm this new mechanism by illustrating that even simplest antiferromagnets, i.e. MnO, can display a magnetically induced ferroelectricity under biaxial strain.

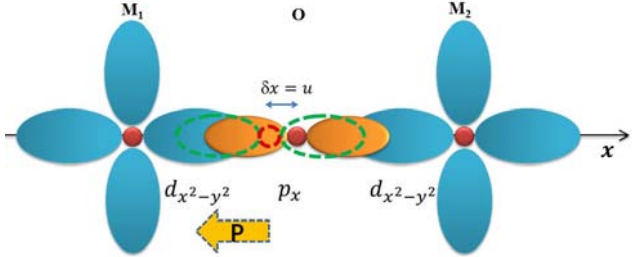

**Figure 1 | Illustration of ferroelectricity induced by indirect magnetic exchange interaction between anion-mediated magnetic cations.** Here $M_1$ and $M_2$ are magnetic cations, O is oxygen anion. When O atom, which originally sits in the middle of $M_1$ and $M_2$ ions, is shifted along the $M_{1(2)}$-O bond direction ($x$) by a small displacement $u$, the magnetic exchange interactions of the system will increase and may support the O off-center movement. An electric dipole is then formed, as shown by an arrow in the picture.

To study the interatomic magnetic exchange interaction and the associated magnetic ordering energy, we consider a three atoms case, where a diamagnetic ion such as oxygen ion sits between two transition metal ions, as shown in Fig.1. As revealed by Sergienlo and Dagotto[13], DMI provides a driving force for the oxygen ion to shift perpendicularly to the spin chain. Due to the small strength, however, the energy gain from DMI is usually less than the ordinary elastic energy, consequently only a few compounds show magnetically induced ferroelectricity[6]. Here, we consider the effect of longitudinal displacement of diamagnetic ions (shown by the dot line in Fig.1) on the magnetic ordering energy. As is well known, in the above case, the distance between magnetic ions usually is much larger than the radii of $d/f$ orbital which carry magnetic moments, thus the direct exchange is negligible, and the hybridization between magnetic and diamagnetic ions is essential for the indirect magnetic exchange coupling. Therefore, we first discuss the hopping integrals in this system. For simplicity, here we only show the case of one orbital per ion, and it is easy to generalize our results to multi-orbital cases. The kinetic energy then can be written as:

$$H_{kin} = \sum_\sigma \varepsilon_d c^+_{d_1\sigma} c_{d_1\sigma} + \sum_\sigma \varepsilon_o c^+_{o\sigma} c_{o\sigma} + \sum_\sigma \varepsilon_d c^+_{d_2\sigma} c_{d_2\sigma}$$
$$+ t_1 \sum_\sigma (c^+_{d_1} c_o + h.c.) + t_2 \sum_\sigma (c^+_{d_2} c_o + h.c.) \quad (1)$$

where the operator $c^+_{d_1\sigma}(c_{d_1\sigma})$, $c^+_{d_2\sigma}(c_{d_2\sigma})$, $c^+_{o\sigma}(c_{o\sigma})$ creates (annihilates) a spin $\sigma$ electron at $M_1$, $M_2$ and O site, respectively. $\varepsilon_d$ and $\varepsilon_o$ are the on-site energies for $M_1/M_2$ and O site. $t_1/t_2$ are the hopping integrals between $M_1/M_2$ and O. Based on the standard Schrieffer-Wolff transformation[22], we can eliminate the O orbital and obtain the effective hopping integral between $M_1$ and $M_2$ site:

$$t_{eff} \propto \frac{t_1 t_2}{\varepsilon_d - \varepsilon_o} \quad (2)$$

It is well known that the hopping integral is inversely proportional to the bond-length: $t_1 \sim r_1^{-y}$, $t_2 \sim r_2^{-y}$, where $r_{1(2)}$ is the bond length of $M_{1(2)}$-O, $y$ is a positive value and strongly depends on both the bond type and the participated orbital[23]. A longitudinal displacement of O ion $u$ will then change the effective hopping to

$$t_{eff} \propto \frac{1}{\varepsilon_d - \varepsilon_0} \frac{1}{(r+u)^y (r-u)^y} = C \frac{1}{(r+u)^y (r-u)^y} \quad (3)$$

where C is a parameter and not sensitive to $u$, $r$ is the distance between magnetic ion and center O-site. It is interesting to notice that regardless the parameter C and the superscript $y$ in Eq. (3), a displacement $u$ which breaks inversion symmetry always increases the effective hopping between the magnetic ions.

To see the effect of $u$ on the magnetic ordering energy, we take the superexchange, which is one of the most common mechanisms in magnetic insulators, as an example. It is well known that for superexchange the interatomic exchange interaction can be written as[24]:

$$J = -\frac{4 t_{eff}^2}{U} \quad (4)$$

where $U$ is the Coulomb interaction for the magnetic orbital. Thus, an off-center displacement of O ion can enhance the interatomic exchange interaction. Usually the magnetic ordering energy has the form of $-\sum_{ij} J_{ij} S_i \cdot S_j$, thus for the AFM case, the energy gain up to the second order of the longitudinal displacement of O ions $u$ is

$$\Delta E \propto -\frac{1}{(r+u)^{2y}(r-u)^{2y}} + \frac{1}{r^{4y}} \approx -\frac{2y}{r^{4y+2}} u^2 \quad (5)$$

Thus we prove that regardless the exact formula for the hopping dependence on distance, an off-center distortion can definitely lower the total energy by an amount $\sim u^2$.

Above we have discussed the superexchange in one-band case. For multi-band superexchange and even double exchange mechanism[25], the exchange coupling $J$ is also proportional to the effective hopping, just the relationship between $J$ and $t_{eff}$ in these mechanisms may not be as simple as shown in Eq. (4). Keeping in mind that the effective hopping is a function of $1/(r^2 - u^2)$, an off-center distortion will therefore always lower the magnetic energy. Moreover, regardless the specific form of $J$ ($t_{eff}$), the corresponding energy gain is still $\sim u^2$, as in the case of the single band superexchange. This is a nontrivial conclusion, since now we see that both the magnetic exchange energy change and the ordinary elastic energy change ($\sim Ku^2/2$) are of the second order of the off-center displacement. Therefore, when the anion located between magnetic ions is shifted away from the center-symmetric point, the increase in elastic energy may be compensated by the decrease of magnetic exchange energy, consequently, forming an electric dipole and resulting in ferroelectricity, as Fig. 1 shows.

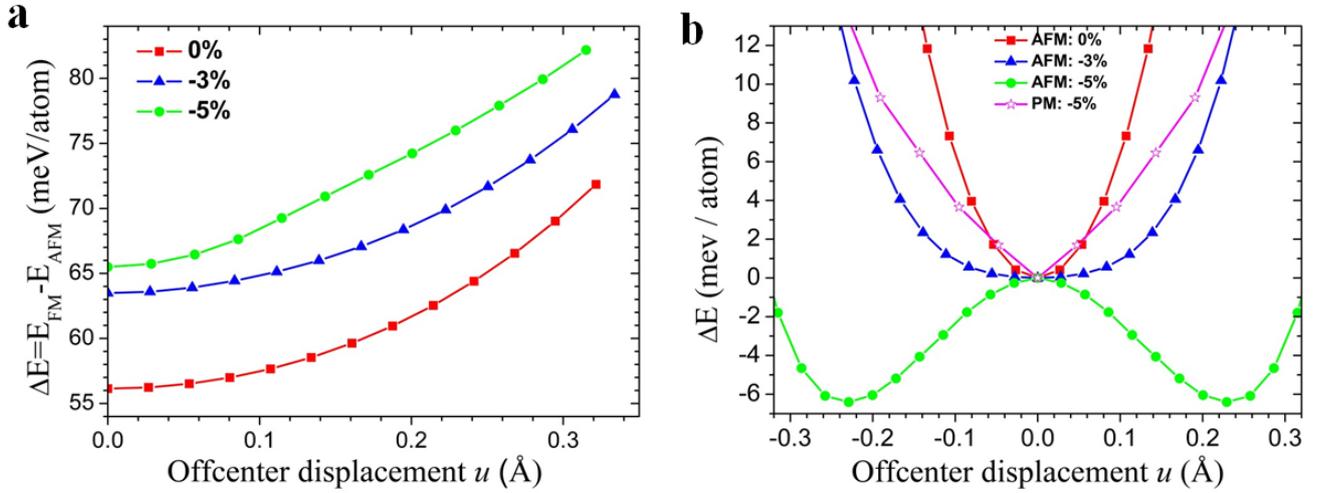

**Figure 2 | Energy difference as a function of O offcenter displacement. a**, Calculated energy differences between zero-temperature AFM and FM state of MnO for a series of O off-center displacement at different biaxial strains: no strain (red square), -3% (blue triangle), -5% (green circle). **b**, Calculated lattice instabilities of MnO under different biaxial strains at zero-temperature AFM and high-temperature PM states (magenta open star). Zero-temperature results come from DFT+$U$ calculations with $U$=3.0 eV, and high-temperature result is from LDA+DMFT simulation with $U$=3.0 eV and T=300 K.

The ferroelectricity is long believed to originate from a delicate balance between the short-range forces favoring the undistorted paraelectric structure and the long range Coulomb interactions favoring the ferroelectric phase[26,27]. Now we see that indirect magnetic exchange, which is short-range in nature, provide another driving force for the off-center atomic motion. To numerically prove the above conclusion, we then carried out a series of first-principles calculations (see Methods). For the purpose of avoiding the complication due to complex magnetic order and achieving an unambiguous result, we choose the simple insulator MnO with rock-salt structure, which is AFM at low temperature ($\sim$ 120 K) and paramagnetic (PM) under normal conditions, as the demonstrating system.

Although cannot deal with the high-temperature PM state of MnO, DFT+$U$ scheme is adequate for the zero-temperature magnetically ordered insulating state[28]. Thus we utilize the DFT+$U$ method to check whether a reasonable biaxial strain can induce ferroelectricity in the ground state of MnO. Here, we concern the off-center (001)-direction motion of O ions. Our numerical results show that regardless the values of $U$ and O-displacement, the ground state of MnO is always AFM. One important parameter relevant to the magnetic ordering energy is the energy difference between AFM and FM state ($\Delta E=E_{FM}-E_{AFM}$). For $U$ = 3 eV[29], the results are shown in Fig. 2a. It is clear from the figure that $\Delta E$ enhances with increasing O displacement, which directly supports our conclusion: off-center motion of the anion between two magnetic cations would enhance the magnetic exchange interaction. As clearly shown in Fig.2b, at strain-free state (0%), the curve for total energy change with the O off-center displacement is parabolic, which indicates a paraelectric state. When the compressive strain applies, the bottom of the curve becomes flat. Then, at a reasonable strain, e. g., -3%, ferroelectricity is induced in the ground state of MnO. At even larger compressive strain (-5%), the energy curve is clearly in the shape of double-well — the symbol of ferroelectricity. The calculated polarization is about 0.38 C/m$^2$, which is comparative to that of typical perovskite ferroelectrics. Furthermore, our numerical results show that enlarging $U$ will suppress the magnetic ordering energy, and consequently the critical strain for the occurrence of ferroelectricity will increase. This is expected from our theory [see Eq. (4)], and again demonstrates the importance of magnetic ordering energy for the ferroelectricity.

To confirm that the above ferroelectric instability is of magnetic origin instead of other mechanism[30], LDA+DMFT calculations are then carried out. We use the highly accurate continue time quantum Monte Carlo (CT-QMC) as the impurity solver[31]. For low temperature, CT-QMC would be very demanding on the computer resource, we thus only consider the high temperature PM phase of MnO by LDA+DMFT. All of our calculations, regardless the temperature (T=200, 300 and 400 K) and $U$, give the same qualitative conclusion: losing the magnetic ordering will suppress the off-center displacement, and even a large strain (-5%) can no longer induce the ferroelectric instability for PM phase (see line with star symbols in Fig. 2b, which is the result of $T$=300 K and $U$=3.0 eV). Thus we unambiguously demonstrate that here the magnetic interaction is essential for the onset of ferroelectricity.

Speaking solely from the point of view of magnetic exchange interaction, antiferroelectric phases are also possible states to lower the magnetic energy of the system. We show here that these states are not energetically favored by considering the electrostatic energy. This is confirmed by our phonon frequency calculation on the paraelectric phase of a 2×2×2 supercell of MnO (64 atoms) under different in-plane strains. We find that only the $A_u$ mode (vibration of Mn and O ion along z-axis) has been gradually softened when the compressive epitaxial strain increases. No lattice instability corresponds to an antiferroelectric phase. Therefore, the off-center O movement will result in ferroelectricity instead of antiferroelectricity.

Our finding, that magnetic exchange interaction could induce ferroelectricity even in simple magnetic oxide insulators, is unexpected and exciting. As shown in the above analysis, the bonus coming with the induced ferroelectricity would be the enhancing of the magnetic transition temperature. In addition, this mechanism is ubiquitous, i.e., not confined in antiferromagnets. Therefore various magnetic systems could be potential multiferroics.

In summary, we have confirmed both analytically and numerically that indirect magnetic exchange, contrary to what people previously thought, favors ferroelectricity even in collinear magnetic systems. Our research provides a new way to explain the coexistence of ferroelectricity and magnetism and might be useful to the search of novel multiferroics suitable of practical use.

**Methods**

For the zero-temperature magnetically ordered insulating state, we use the projector augmented wave (PAW) method implemented in the Vienna *Ab-Initio* Simulation Package (VASP)[32]. The exchange-correlation potential is treated in the generalized gradient approximation (GGA). We use the energy cut-off of 500 eV for the plane wave expansion of the PAWs and a 10 x 10 x 10 Monkhorst-Pack grid for *k*-point sampling in the self-consistent calculations. We vary the onsite Coulomb energy $U$ from 1 to 6 eV and confirm its value do not change our qualitative conclusions. The epitaxial strain is defined as $(a - a_0)/a_0$, where $a$ is the in-plane lattice parameter and $a_0$ is the theoretical equilibrium lattice constant in cubic symmetry. The out-of-plane lattice parameter $c$ is optimized at every strain. The Berry phase technique is used to calculate ferroelectric polarizations.

We use the LDA+DMFT method[28,31] to calculate the high temperature PM state. We use continue time quantum Monte Carlo as the impurity solver[31], and crossing check our results by the non-crossing approximation. Calculations are fully self-consistent in charge density, chemical potential, impurity level and total energy. The line with star symbol in Fig.2b shows the LDA+DMFT results for -5% strain with temperature equal to 300K and $U$ equal to 3.0 eV.

**Acknowledgements** XW thanks Kristjan Haule for useful discussion. The work was supported by the National Key Project for Basic Research of China (Grant no. 2011CB922101 and 2010CB923404), NSFC (Grant No. 91122035, 61125403, 50832003); PCSIRT, NCET, Shanghai ShuGuang. Computations were performed at the ECNU computing center. S.Y.S was supported by DOE Computational Material Science Network (CMSN) and DOE SciDAC Grant No. SE-FC02-06ER25793.

Correspondence and requests for materials should be addressed to X. W or C.G. D.